\documentclass{desyproc}
\usepackage{titlesec}
\titlespacing{\section}{0pt}{7pt}{5pt}
\def\jpsi{$\mathrm{J}/\psi${}}
\def\psip{$\psi(\mathrm{2S})$}
\def\ro{$\rho^0$}
\def\etal{{\it et~al.}}
\begin{document}
\title{ALICE Results on Vector Meson \mbox{Photoproduction} in Ultra-peripheral p--Pb and Pb--Pb Collisions}
\author{{\slshape Evgeny Kryshen for the ALICE Collaboration}\\[1ex]
CERN, 1211 Geneva 23, Switzerland }
\contribID{198}
\confID{8648}
\desyproc{DESY-PROC-2014-04}
\acronym{PANIC14}
\doi
\maketitle
\begin{abstract}
Lead nuclei, accelerated at the LHC, are sources of strong electromagnetic fields that can be used to measure photon-induced interactions in a new kinematic regime. These interactions can be studied in ultra-peripheral p--Pb and Pb--Pb collisions where impact parameters are larger than the sum of the nuclear radii and hadronic interactions are strongly suppressed. Heavy quarkonium photoproduction is of particular interest since it is sensitive to the gluon distribution in the target. The ALICE Collaboration has studied \jpsi\ and \psip\ photoproduction in ultra-peripheral Pb--Pb collisions and exclusive \jpsi\ photoproduction off protons in ultra-peripheral p--Pb collisions at the LHC. Implications for the study of gluon density distributions and nuclear gluon shadowing are discussed. Recent ALICE results on  \ro\ photoproduction are also presented.
\end{abstract}

\section{Introduction}

Lead nuclei, accelerated at the LHC, are sources of strong electromagnetic fields, which are equivalent to a flux of quasi-real photons, thus p--Pb and Pb--Pb collisions can be used to measure $\gamma$p, $\gamma$Pb and $\gamma\gamma$ interactions in a new kinematic regime. These interactions are usually studied in ultra-peripheral collisions (UPC), characterised by impact parameters larger than the sum of the radii of the incoming hadrons, in which hadronic interactions are strongly suppressed. Heavy quarkonium photoproduction is of particular interest since, in leading order perturbative QCD, its cross section is proportional to the squared gluon density of the target. LHC kinematics corresponds to Bjorken-$x$ ranging from $x \sim 10^{-2}$ down to $x \sim 10^{-5}$, while the heavy-quark mass requires a virtuality $Q^2$ larger than a few GeV$^2$, hence introducing a hard scale. Thus quarkonium photoproduction off protons in p--Pb UPC can be used to probe the behaviour of the gluon density at low $x$ and to search for gluon saturation in the proton. On the other hand, quarkonium photoproduction in \mbox{Pb--Pb} UPC provides a direct tool to study nuclear gluon shadowing effects, which are poorly known and play a crucial role in the initial stages of heavy-ion collisions. Light vector meson photoproduction measurements would also help to shed light on underlying photoproduction mechanisms at a soft scale.

The ALICE experiment measured \jpsi, \psip\ and \ro\ photoproduction in Pb--Pb UPC at~$\sqrt{s_{\mathrm{NN}}} = 2.76\ {\rm TeV}$ and exclusive J$/\psi$ photoproduction off protons in p--Pb UPC  at~$\sqrt{s_{\mathrm{NN}}} = 5.02\ {\rm TeV}$~\cite{alice-forward,alice-central,alice-pAforward}. These results are briefly reviewed in the following sections.

\section{\jpsi, \psip\ and \ro\ photoproduction in Pb--Pb collisions} 

The ALICE detector~\cite{alice-review} consists of a central barrel covering the pseudorapidity range $|\eta| < 0.9$ and a  muon spectrometer in the forward direction. Central barrel detectors, relevant for UPC measurements, include an Inner Tracking System (ITS), a Time Projection Chamber (TPC) and a Time-Of-Flight detector (TOF). The muon spectrometer consists of a set of absorbers, a dipole magnet, five tracking and two trigger stations used to detect muons in the range $-4 < \eta < -2.5$. The VZERO-A ($2.8 < \eta < 5.1$) and the VZERO-C ($-3.7 < \eta < -1.7$) scintillator arrays are used for triggering and multiplicity measurements. The Zero-Degree Calorimeters (ZDC), located at $\pm 114$~m from interaction point, are used to detect neutrons in the very forward regions. 

ALICE measured \jpsi\ photoproduction in ultra-peripheral Pb--Pb collisions at forward rapidity in the dimuon channel~\cite{alice-forward} and at central rapidity both in the dimuon and dielectron channels~\cite{alice-central}. The forward UPC trigger required a single muon with $p_{\rm T} > 1$\ GeV/$c$ in the muon spectrometer, at least one hit in VZERO-C and a veto on \mbox{VZERO-A} activity. For the measurement at central rapidity, the trigger required two back-to-back hits in TOF, two hits in a Silicon Pixel Detector (SPD, two innermost ITS layers) and vetoes on both VZERO detectors. 
Events with only two unlike sign dileptons and a neutron ZDC signal below 6 TeV were then selected in the offline analysis. The energy deposition in the TPC was used to separate dielectron and dimuon channels at mid-rapidity.

\begin{figure}[b!]
\vskip -9pt
\centerline{
\includegraphics[width=0.445\textwidth,trim= 0 2 35 35, clip=true]{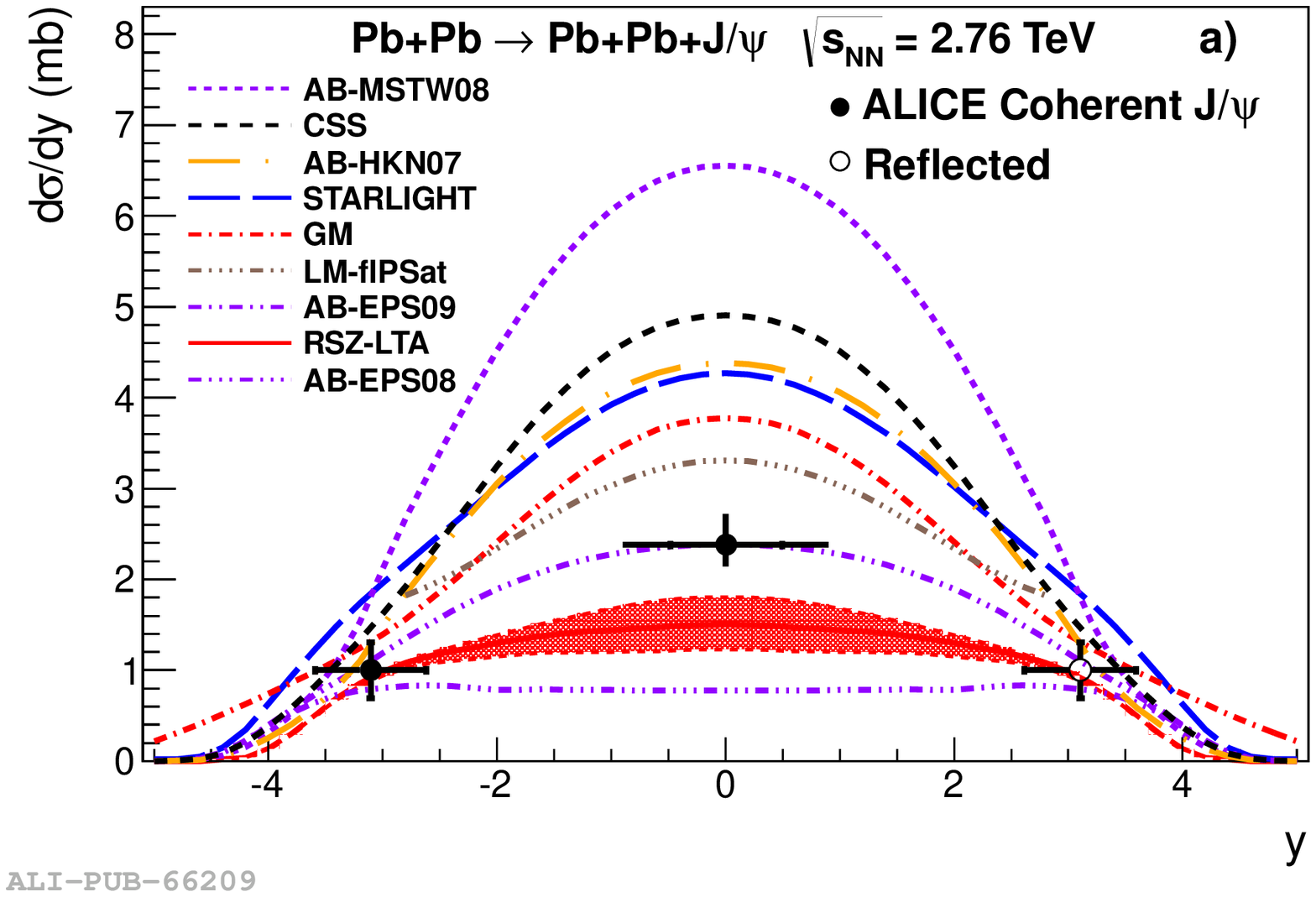}
\includegraphics[width=0.555\textwidth,trim= 3 2 20 12, clip=true]{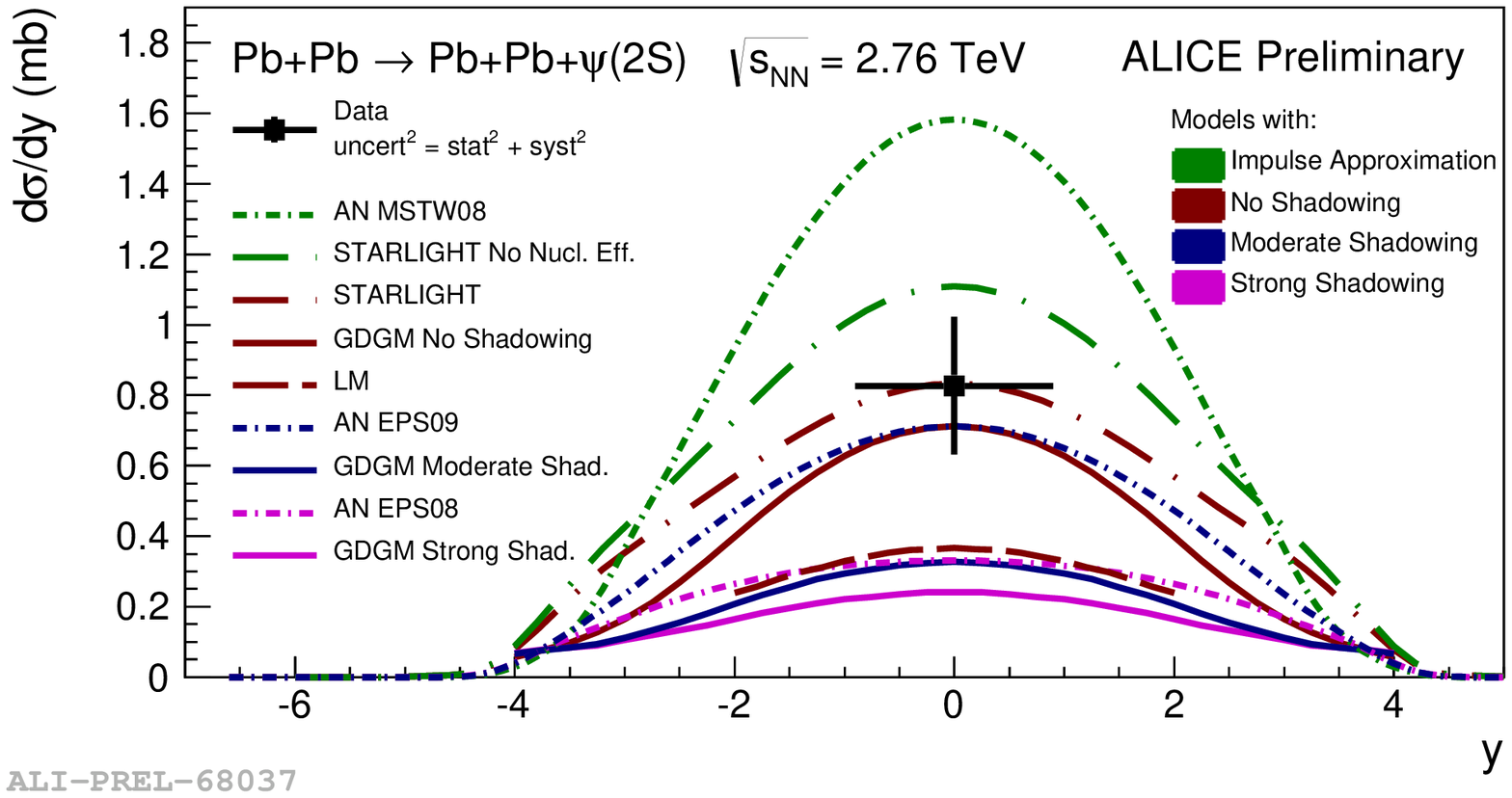}
}
\vskip -3pt
\caption{ALICE results on coherent \jpsi\  (left) and \psip\ (right) photoproduction cross section in Pb--Pb collisions at $\sqrt{s_{\mathrm{NN}}} = 2.76 {\rm\ TeV}$ in comparison with model predictions~\cite{alice-forward,alice-central}.}
\label{Fig:1}
\end{figure}

The reconstructed \jpsi\ signal includes contributions from coherent and incoherent \mbox{production} mechanisms. Coherent \jpsi\ photoproduction, when a photon interacts coherently with the~whole nucleus, is characterized by a narrow transverse momentum distribution with~$\langle p_{\rm T} \rangle \sim 60$~MeV$/c$. In the incoherent case the photon couples to a single nucleon so that the $p_{\rm T}$ distribution becomes much broader with $\langle p_{\rm T} \rangle \sim 400$~MeV$/c$. The transverse momentum distributions for di\-lep\-tons with an invariant mass around the \jpsi\ mass were fitted with templates \mbox{corresponding} to different production mechanisms. Contributions from continuum dilepton production, \mbox{feed-down} from \psip\ decays and a possible contamination from hadronically produced \jpsi{}'s were also taken into account in the fits. The results on the coherent \jpsi{} photoproduction cross section are compared with various model calculations in Fig.~\ref{Fig:1} (left). The best agreement was found for the model~\cite{ab}, which incorporates gluon shadowing according to EPS09 global fits~\cite{Eps09}. A similar conclusion was obtained in Ref.~\cite{guzey} where the gluon shadowing factor $R_g(x\sim 10^{-3}, Q^2 \sim 2.4\ {\rm GeV}^2) = 0.61^{+0.05}_{-0.04}$ was extracted from the ALICE measurement at mid-rapidity.

ALICE also measured coherent \psip\ cross section at mid-rapidity via the dilepton ($l^{+} l^{-}$) decay and in the channel  $ \psi(\mathrm{2S}) \rightarrow {\rm J}/\psi + \pi^{+} \pi^{-}$ followed by ${\mathrm J}/\psi \rightarrow l^{+} l^{-}$ decay. The measured cross section, shown in Fig.~\ref{Fig:1} (right), disfavours models with no nuclear effects and those with strong gluon shadowing, however different predictions rely on different reference $\gamma+{}{\rm p} \to \psi(\mathrm{2S})+{}{\rm p}$ cross sections, thus preventing stronger conclusions. Many uncertainties on the measurement and on the $\gamma$p reference cancel in the ratio of the coherent \psip\ and \jpsi\ cross sections. The measured ratio $\sigma_{\psi(\mathrm{2S})}^{\rm coh}/\sigma_{{\rm J}/\psi}^{\rm coh} = 0.344_{-0.074}^{+0.076}$ appears to be a factor two larger than in $\gamma$p measurements at HERA~\cite{psip} indicating that nuclear effects may affect differently 1S and 2S charmonium states.

Measurement of the coherent \ro\ photoproduction at the LHC is important for verification of \ro\ photoproduction models, which differ by factor two in the predicted cross sections~\cite{sl,gm,rho}. ALICE measured coherent \ro\ photoproduction cross section in the $\pi^+\pi^-$ channel at mid-rapidity in ultra-peripheral Pb--Pb collisions at $\sqrt{s_{\mathrm{NN}}} = 2.76$\ TeV. ALICE results, shown in Fig.~\ref{Fig:2} (left), disfavour the standard Glauber approach (GDL1 curve)~\cite{rho}, but appear to be in agreement with STARLIGHT, which is also based on the Glauber formalism, but neglects the elastic part of the total $\rho N$ cross section~\cite{sl}. It is worth noting that a similar trend has been already revealed at lower energies by the STAR experiment~\cite{star}. The ALICE measurement is also consistent with the GM model~\cite{gm} based on the colour-dipole approach and the Color Glass Condensate formalism. 

\section{Exclusive \jpsi\ photoproduction in p--Pb collisions} 

The large photon flux produced by the lead nucleus in p--Pb collisions at the LHC offers a possibility to measure exclusive \jpsi\  photoproduction off protons and to probe the gluon density distribution in the proton in a new kinematic regime. \jpsi\ photoproduction has been previously studied at HERA at $\gamma {\rm p}$ centre-of-mass energies $W_{\gamma {\rm p}}$ ranging from 20 to 305 GeV~\cite{hera}. HERA cross sections are well described by a power law $\sigma(W_{\gamma {\rm p}}) \sim W_{\gamma {\rm p}}^{\delta}$, reflecting the fact that the underlying gluon distribution follows a power law in $x$ down to $x\sim 10^{-4}$. A deviation from the power law for  $\sigma(W_{\gamma {\rm p}})$ at higher energies could indicate a change in the evolution of the gluon density function at lower $x$, as expected at the onset of saturation.

\begin{figure}[b!]
\vskip -9pt
\centerline{
\includegraphics[width=0.45\textwidth,trim= 3 2 55 34, clip=true]{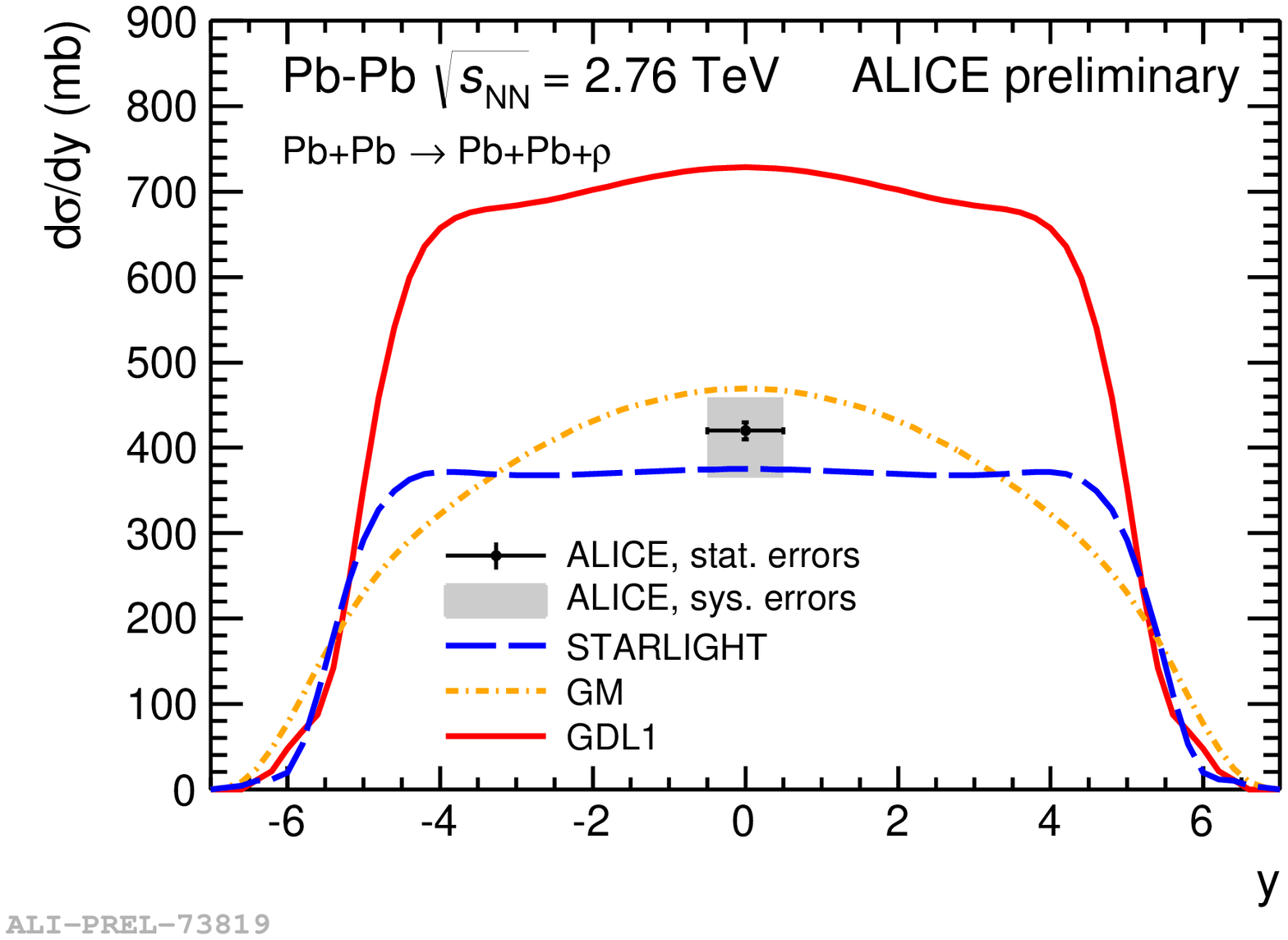}
\includegraphics[width=0.50\textwidth,trim= 0 2 0 5, clip=true]{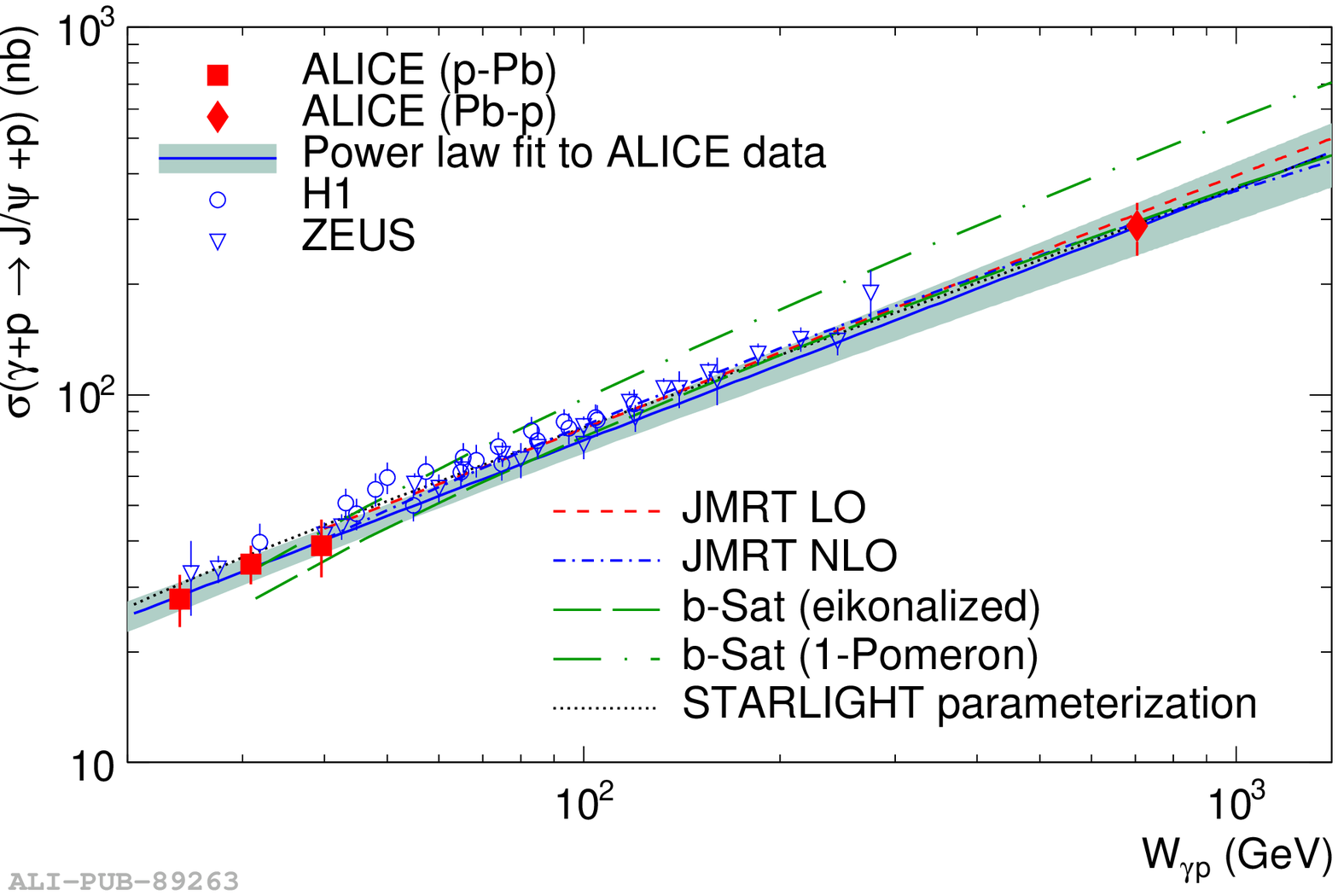}
}
\vskip -5pt
\caption{Left: ALICE results on coherent $\rho^0$ photoproduction cross section in Pb--Pb collisions at $\sqrt{s_{\mathrm{NN}}} = 2.76 {\rm\ TeV}$. Right: exclusive \jpsi\ photoproduction cross section off protons measured by ALICE in comparison with HERA data and model predictions~\cite{alice-pAforward}. }
\label{Fig:2}
\vskip -10pt
\end{figure}

ALICE utilized three options to trigger on \jpsi\ decays in ultra-peripheral p--Pb collisions: dimuon in the muon spectrometer ($-4.0 < \eta < -2.5$), dimuon or dielectron in the central barrel ($|\eta|<0.9$) and intermediate case with a muon in the muon arm and another one in the barrel.  In addition, the LHC provided collisions of protons on lead ions in two configurations: p--Pb (the proton moves towards the muon spectrometer) and Pb--p (the lead ion moves towards~the muon spectrometer). This allowed ALICE to study exclusive \jpsi\ photoproduction in a rapidity range $-4 < y <4$, extend accessible $W_{\gamma {\rm p}}$ energies up to almost 1 TeV and~to probe Bjorken-$x$ down to $x\sim 2 \times 10^{-5}$ where saturation effects might already play an important role.

The first ALICE results on exclusive \jpsi\ photoproduction off protons measured in p--Pb collisions via dimuon channel at forward rapidity were published in~\cite{alice-pAforward}. \jpsi\ decays were reconstructed in $2.5 < y < 4.0$ (p--Pb) and $-3.6 < y < -2.6$ (Pb--p) rapidity intervals, corresponding to $\gamma {\rm p}$ centre-of-mass energies of $21< W_{\gamma {\rm p}} <45$\ GeV and $577< W_{\gamma {\rm p}} <952$\ GeV respectively. Exclusive  \jpsi\ events were selected by vetoing activity in SPD, VZERO-A and ZDC. The remaining non-exclusive \jpsi, e.g. diffractive events accompanied  by the proton dissociation, were subtracted out by fitting $p_{\rm T}$ distributions with templates corresponding to exclusive and non-exclusive event samples. Photoproduction cross sections  $\sigma(W_{\gamma {\rm p}})$ were extracted from the exclusive cross sections $\frac{{\rm d}\sigma}{{\rm d}y}({\rm p}+{\rm Pb} \to {\rm p}+{\rm Pb} + {\rm J}/\psi)$ at the corresponding rapidity intervals by dividing out the photon flux from Pb nuclei. ALICE results on $\sigma(W_{\gamma {\rm p}})$ are shown in Fig.~\ref{Fig:2} (right) in comparison with previous measurements and various theoretical models based on extrapolations of HERA data. ALICE photoproduction cross sections are consistent with a power law with $\delta = 0.68 \pm 0.06$ (stat + syst), similar to the trend obtained from HERA, thus indicating no significant change in the power-law $x$-dependence of the gluon density in  the proton down to $x \sim 2 \times 10^{-5}$.

\section{Conclusions}

ALICE results on the coherent \jpsi\ photoproduction in ultra-peripheral Pb--Pb collisions are in good agreement with models based on the moderate gluon shadowing from the EPS09 global fit. The measured coherent \psip\   photoproduction cross section disfavours models with no nuclear effects and those with strong gluon shadowing. Coherent \ro\ photoproduction in Pb--Pb collisions at the LHC cannot be described in the standard Glauber formalism, but is in agreement with STARLIGHT and the colour-dipole approach. ALICE results on the exclusive \jpsi\ photoproduction off a proton measured in p--Pb collisions indicate no \mbox{significant change in} the power-law $x$-dependence of the gluon density in the proton between HERA and LHC energies. 

\begin{footnotesize}

\end{footnotesize}
\end{document}